\documentclass[aps,prb,twocolumn,showpacs,amsmath,amssymb,superscriptaddress,floatfix]{revtex4-1}
\usepackage{graphicx}
\usepackage{color}
\usepackage{braket}
\usepackage{amsfonts}
\usepackage{MnSymbol}
\usepackage{bm}

\usepackage[dvipsnames]{xcolor}
\usepackage{mathrsfs}
\usepackage{mathtools}
\usepackage{subfigure}

\usepackage[colorlinks = true,linkcolor = blue,urlcolor  = blue,     citecolor = blue,anchorcolor = blue]{hyperref}

\usepackage{bbm}
\usepackage[dvipsnames]{xcolor}

\date{\today}                  % Activate to display a given date or no date

\begin{document}

\title{Topological Josephson vortices at finite voltage bias}

\author{Kiryl Piasotski}
\email[Email: ]{kiryl.piasotski@kit.edu}
\affiliation{Institut f\"ur Theorie der Kondensierten Materie, Karlsruher Institut f\"ur Technologie, 76131 Karlsruhe, Germany}
\affiliation{Institut f\"ur Quanten Materialien und Technologien, Karlsruher Institut f\"ur Technologie, 76021 Karlsruhe, Germany}

\author{Adrian Reich}
\affiliation{Institut f\"ur Theorie der Kondensierten Materie, Karlsruher Institut f\"ur Technologie, 76131 Karlsruhe, Germany}

\author{Alexander Shnirman}
\affiliation{Institut f\"ur Theorie der Kondensierten Materie, Karlsruher Institut f\"ur Technologie, 76131 Karlsruhe, Germany}
\affiliation{Institut f\"ur Quanten Materialien und Technologien, Karlsruher Institut f\"ur Technologie, 76021 Karlsruhe, Germany}

\begin{abstract}
We study the effects of finite voltage bias on Caroli-de Gennes-Matricon (CdGM) states in topological Josephson junctions with a vortex lattice. The voltage drives vortices into steady motion, squeezing the CdGM spectrum due to quasi-relativistic dispersion. A finite voltage range allows well-defined states, but beyond a critical breakdown voltage, the states collapse to zero energy and become sharply localized, marking a dynamical transition. Additionally, finite bias modifies selection rules for CdGM state transitions. Notably, in the steady-state regime, the time-averaged current vanishes, revealing a novel interplay between vortex dynamics and quantum coherence.  
\end{abstract}

\maketitle

\section{Introduction}

Majorana fermions have attracted significant interest in condensed matter physics due to their potential applications in topological quantum computation~\cite{kitaev2001unpaired}. These exotic quasiparticles obeying anyonic statistics are predicted to emerge as zero-energy modes in topological superconductors~\cite{sato2017topological} and their artificial counterparts~\cite{oreg2010helical,lutchyn2010majorana,fu2008superconducting,fu2009josephson,Nadj-PerdePRB2013}, and are commonly thought of as promising candidates for fault-tolerant quantum computing~\cite{kitaev2003fault}. Topological insulator-superconductor (TI-S) junctions provide an ideal setting to explore Majorana physics, leveraging the unique helical surface states of three-dimensional topological insulators~\cite{fu2008superconducting,fu2009josephson}. In particular, the Josephson vortices generated by external transverse magnetic fields, in such junctions can host Majorana zero modes, which have been the focus of extensive theoretical~\cite{potter2013anomalous, Laubscher2024, piasotski2024topological} and experimental~\cite{zhang2022ac,yue2024signatures, park2024corbino} investigations. Understanding the dynamics of these vortices, especially under external perturbations such as magnetic fields or voltage bias, is crucial for advancing our knowledge of topological superconductivity and its applications in quantum information science.

In this work, we investigate the impact of a finite voltage bias applied across a two-dimensional (2D) topological Josephson junction on the spectrum and localization of Caroli-de Gennes-Matricon (CdGM) states. The applied voltage drives a steady motion of Josephson vortices, with their velocity proportional to the voltage bias. This motion induces spectral squeezing of the CdGM states due to their quasi-relativistic dispersion. Remarkably, we find that the CdGM states remain well-defined within a finite range of voltages. Beyond a critical breakdown voltage, the states collapse to zero energy and become sharply localized, marking a novel dynamical transition in the behavior of topological Josephson junctions.

In addition to revealing this voltage-induced spectral transition, we demonstrate that the steady vortex motion gives rise to a distinctive dynamical property: the vanishing of the DC component of the current across the junction. This counterintuitive result highlights the intricate interplay between the voltage bias, vortex dynamics, and spectral characteristics of the system, offering new insights into the transport properties of topological Josephson junctions under non-equilibrium conditions.

These findings not only enhance our understanding of the fundamental physics governing voltage-driven dynamics in topological systems but also suggest new directions for experimental exploration. The predicted critical behavior and vanishing DC current provide testable signatures that could pave the way for uncovering new dynamical phenomena in 2D topological Josephson junctions.

\begin{figure}
\centering
                \includegraphics[scale=0.14]{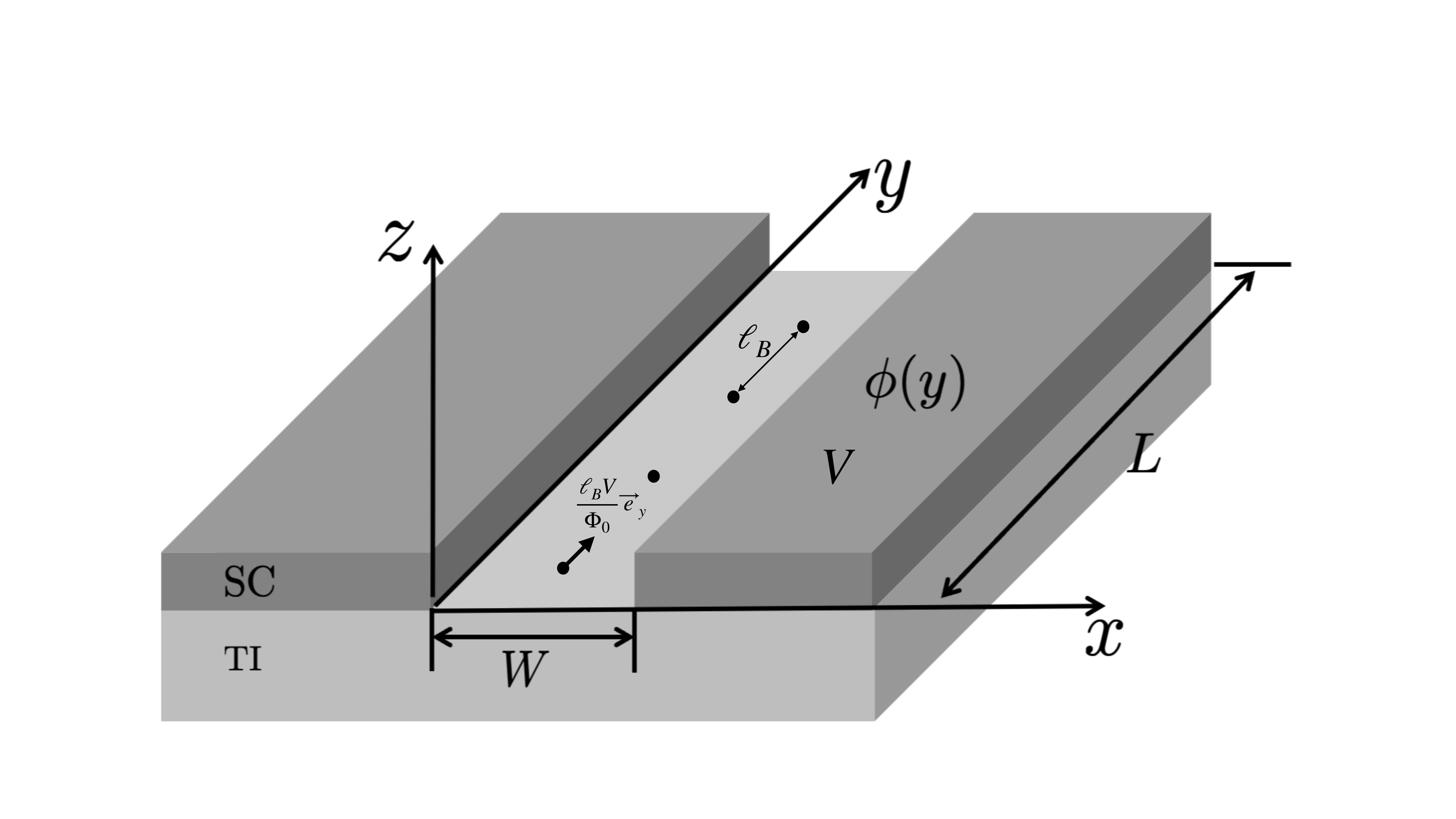}
                \caption{A two-dimensional topological Josephson junction on the surface of a 3D topological insulator. An externally applied magnetic field induces a lattice of Josephson vortices in the junction's tunneling region. The application of an external voltage sets the vortex lattice into steady motion along the junction, with the motion rate determined by the voltage bias $V$ and the superconducting magnetic length $\ell_B$.}
                \label{SKETCH}
\end{figure}

\section{Equations of motion}
\subsection{Microscopic model}
We consider a planar Josephson junction on the surface of a 3D topological insulator (TI), as illustrated in Fig. \ref{SKETCH}. The essential low-energy physics of this setup is captured by the following Bogoliubov-de Gennes (BdG) Hamiltonian: 
\begin{align}
    H(t) = \frac{1}{2} \int d^{(2)}\bold{r} \, \Psi^{\dagger}(\bold{r}) h(t) \Psi(\bold{r}),
\end{align}
where the Nambu spinor is defined as  
\begin{align}
    \Psi(\bold{r}) = \begin{pmatrix} \psi_{\uparrow}(\bold{r}) \\ \psi_{\downarrow}(\bold{r}) \\ \psi_{\downarrow}^{\dagger}(\bold{r}) \\ -\psi_{\uparrow}^{\dagger}(\bold{r}) \end{pmatrix},  
\end{align}
and the single-particle Hamiltonian \( h(t) \) is given by 
\begin{align}
\nonumber
    h(t) =& v_F \, \bm{\sigma} \cdot \left( \bold{p} + \frac{e}{c} \tau_z \bold{A}(\bold{r}) \right) \tau_z  
- \mu(\bold{r}) \tau_z \\
&+ \left( \Delta(\bold{r}, t) \tau_{+} + \text{h.c.} \right). 
\end{align}

Here, \(\bold{r} = (x, y)\), \(\bold{p} = -i(\partial_x, \partial_y)\), \(\bm{\sigma} = (\sigma_x, \sigma_y)\), and \(\tau_i\) (\(i = x, y, z, +\)) are Pauli matrices acting in spin and particle-hole spaces, respectively. The parameters are as follows: \(v_F\) is the Fermi velocity of the linearly dispersing surface states of the 3D TI; \(\mu(\bold{r})\), \(\bold{A}(\bold{r})\), and \(\Delta(\bold{r}, t)\) represent the chemical potential, magnetic vector potential, and proximity-induced superconducting pairing amplitude, respectively. Throughout, we use \(e > 0\).

The proximity-induced superconducting gap \(\Delta(\bold{r}, t)\) has a step-like spatial profile:  
\begin{align}
    \Delta(\bold{r}, t) = \Delta \Theta(-x) + e^{i\varphi(y, t)} \Delta \Theta(x - W),
\end{align}
where the space-time-dependent phase difference \(\varphi(y, t)\) is given by  
\begin{align}
    \varphi(y, t) = \frac{2\pi y}{\ell_B} - \frac{2\pi}{\Phi_0} \Phi(t).
\end{align}
The term \(\frac{2\pi y}{\ell_B}\) originates from the external magnetic field in the Landau gauge, while the time-dependent phase \(\frac{2\pi}{\Phi_0} \Phi(t)\) arises due to the external voltage \( \Phi(t) = \int^t dt' V(t')\). Instead of using a flux (current) drive, as discussed in Refs. [\onlinecite{Devoret2020, Devoret2021, Houzet2024, piasotski2024theory}], an alternative approach could involve a gate drive, similar to the methods explored in Refs. [\onlinecite{Heck2023, Nazarov2010}].

To incorporate the renormalization of the chemical potential due to superconducting proximity and external drive, we assume a step-like profile for the chemical potential:  
\begin{align}
    \mu(x) &= \mu_S [\Theta(-x) + \Theta(x - W)] + \mu_N \Theta(x) \Theta(W - x).
\end{align}
In this model, we adopt the Andreev limit, assuming that \(\mu_S\) is the largest energy scale.

\subsection{Effective single-vortex dynamics}

In Refs.~[\onlinecite{potter2013anomalous, Laubscher2024, piasotski2024topological}], it was demonstrated that the effective low-energy behavior of a single topological Josephson vortex can be captured by the following Nambu Hamiltonian:
\begin{align}
    h_{\text{eff}}(t) = v\sigma_y p_y + \alpha y\sigma_x - \frac{\ell_B \alpha}{\Phi_0} \sigma_x \Phi(t),
\end{align}
where \(v\) and \(\alpha\) are related to the microscopic parameters of the model as:  
\begin{align}
v =& \frac{v_F}{1 + W/\xi} \frac{\Delta}{\mu_N} \sin\left(\frac{\mu_N}{\Delta} \frac{W}{\xi}\right) + \mathcal{O}(\Delta/\mu_S),\\
    \alpha =& \frac{\Delta}{2} \frac{1}{1 + W/\xi} \frac{2\pi}{\ell_B}.
\end{align}

At zero voltage bias (\(\Phi(t) = 0\)), the eigenstates and eigenenergies of \(h_{\text{eff}}\) were previously determined in Refs. [\onlinecite{Laubscher2024, piasotski2024topological}]:  
\begin{align}
\chi_0(y) =& i \begin{pmatrix} 1 \\ 0 \end{pmatrix} \frac{\phi_0(y/\lambda_B)}{\sqrt{\lambda_B}}, \quad E_0 = 0,\\
\chi_{n, \tau}(y) =& \frac{1}{\sqrt{2}} \left( \begin{pmatrix} 1 \\ 0 \end{pmatrix} \frac{\phi_n(y/\lambda_B)}{\sqrt{\lambda_B}} + \tau \begin{pmatrix} 0 \\ 1 \end{pmatrix} \frac{\phi_{n-1}(y/\lambda_B)}{\sqrt{\lambda_B}} \right),\\
    E_{n, \tau} =& \tau \omega_B \sqrt{n}, \quad \tau = \pm, \ n \in \mathbb{N},
\end{align}
where \(\phi_n(x) = \frac{e^{-x^2/2}}{\sqrt{2^n n! \sqrt{\pi}}} H_n(x)\) are the harmonic oscillator eigenfunctions. The localization length \(\lambda_B\) and the fundamental frequency \(\omega_B\) are given by:  
\begin{align}
\lambda_B =& \left|\frac{\xi \ell_B}{\pi} \frac{\Delta}{\mu_N} \sin\left(\frac{\mu_N}{\Delta} \frac{W}{\xi}\right)\right|^{1/2},\\
    \omega_B =& \frac{\Delta}{1 + W/\xi} \left|\frac{2\pi \xi}{\ell_B} \frac{\Delta}{\mu_N} \sin\left(\frac{\mu_N}{\Delta} \frac{W}{\xi}\right)\right|^{1/2}.
\end{align}

An important characteristic of these vortex states is the sequence of current selection rules identified in Refs. [\onlinecite{Laubscher2024, piasotski2024topological}]. In this low-energy theory, the effective current vertex is given by:  
\begin{align}
    j = \frac{\ell_B \alpha}{\Phi_0} \sigma_x,
\end{align}
which yields the following matrix elements for transitions between states:  
\begin{align}
    I_{n, n+1} =& I_{n, -(n+1)} = -I_{n+1, -n} = -I_{-(n+1), -n} = \frac{1}{2} \bar{I},\\
I_{0, 1} =& I_{0, -1} = -I_{-1, 0} = -I_{1, 0} = \frac{1}{i\sqrt{2}} \bar{I},
\end{align}
where \(\bar{I} = \frac{\ell_B \alpha}{\Phi_0}\).  

Having established the structure of the Caroli-de Gennes-Matricon (CdGM) states for a topological Josephson vortex, we now turn to their behavior under the influence of an applied external voltage.

Let us now focus on the case of a constant voltage bias, where \(\Phi(t) = Vt\). In this case, the effective Hamiltonian becomes:

\begin{align}
    h_{\text{eff}}(t) = e^{-i \frac{\ell_{B} V t}{\Phi_0} p_y} h_{\text{eff}} e^{i \frac{\ell_{B} V t}{\Phi_0} p_y},
\end{align}

where \(h_{\text{eff}} = v \sigma_y p_y + \alpha y \sigma_x\) is the single-vortex Hamiltonian in the voltage-free state. By performing the gauge transformation \(\psi(y,t) = e^{-i \frac{\ell_{B} V t}{\Phi_0} p_y} \tilde{\psi}(y,t)\), we obtain an autonomous Schrödinger equation for the transformed wavefunction \(\tilde{\psi}(y,t)\):

\begin{align}
    i \frac{d}{dt} \tilde{\psi}(y,t) = \tilde{h}_{\text{eff}} \tilde{\psi}(y,t),
\end{align}

where the transformed Hamiltonian is given by

\begin{align}
    \tilde{h}_{\text{eff}} = h_{\text{eff}} - \frac{\ell_{B} V}{\Phi_0} p_y = v \left[\sigma_y - \beta \sigma_0\right] p_y + \alpha y \sigma_x,
\end{align}
with $v\beta = \frac{\ell_{B} V}{\Phi_0}$.

This transformation allows us to associate a conserved quasi-energy in the new reference frame, \(\tilde{\psi}(y,t) = e^{-i \tilde{E} t} \tilde{\psi}_{\tilde{E}}(y)\), and study the eigenvalue problem for \(\tilde{h}_{\text{eff}}\):

\begin{align}
    \tilde{h}_{\text{eff}} \tilde{\psi}_{\tilde{E}}(y) = \tilde{E} \tilde{\psi}_{\tilde{E}}(y).
\end{align}

We define the matrix \(\mathcal{U}\) as
\begin{align}
    \mathcal{U} = \frac{1}{\sqrt{2\gamma}} \begin{pmatrix}
\sqrt{\gamma+1} & -i \text{sgn}(\beta) \sqrt{\gamma-1} \\
i \text{sgn}(\beta) \sqrt{\gamma-1} & \sqrt{\gamma+1}
\end{pmatrix},
\end{align}

where the Lorentzian factor \(\gamma\) is expressed as
\begin{align}
    \gamma = \frac{1}{\sqrt{1 - \beta^2}}. 
\end{align}

The matrix \(\mathcal{U}\) has the following properties:
\begin{align}
    \mathcal{U}^{\dagger} =& \mathcal{U}, \quad \mathcal{U}^2 = 1 + \beta \sigma_y,\\
    \mathcal{U} \sigma_x \mathcal{U} =& \frac{1}{\gamma} \sigma_x, \quad \mathcal{U} \left[\sigma_y - \beta \sigma_0\right] \mathcal{U} = \frac{1}{\gamma^2} \sigma_y
\end{align}

Next, we perform another transformation:
\begin{align}
    \tilde{\psi}_{\tilde{E}}(y) = e^{i \gamma^2 \beta \tilde{E} y / v} \mathcal{U} \hat{\psi}_{\tilde{E}}(z), \quad z = \gamma^{1/2} y,
\end{align}
which modifies the stationary Schrödinger equation to:
\begin{align}
    \hat{h}_{\text{eff}} \hat{\psi}_{\tilde{E}}(z) = \tilde{E} \hat{\psi}_{\tilde{E}}(z)
\end{align}
with a new Hamiltonian
\begin{align}
    \hat{h}_{\text{eff}} = \frac{1}{\gamma^{3/2}} \left(v \sigma_y p_z + \alpha z \sigma_x\right).
\end{align}
From this, we immediately obtain the eigenvalues
\begin{align}
    \tilde{E}_{n, \tau} = \tau \frac{\omega_B}{\gamma^{3/2}} \sqrt{n},
\end{align}
and the eigenstates
\begin{align}
    \tilde{\psi}_0(y) =& \gamma^{1/4} \mathcal{U} \chi_0(\gamma^{1/2} y),\\
    \tilde{\psi}_{n, \tau}(y) =& \gamma^{1/4} e^{i \gamma^2 \beta \tilde{E}_{n, \tau} y / v} \mathcal{U} \chi_{n, \tau}(\gamma^{1/2} y).
\end{align}

\begin{figure}
\centering
                \includegraphics[scale=0.22]{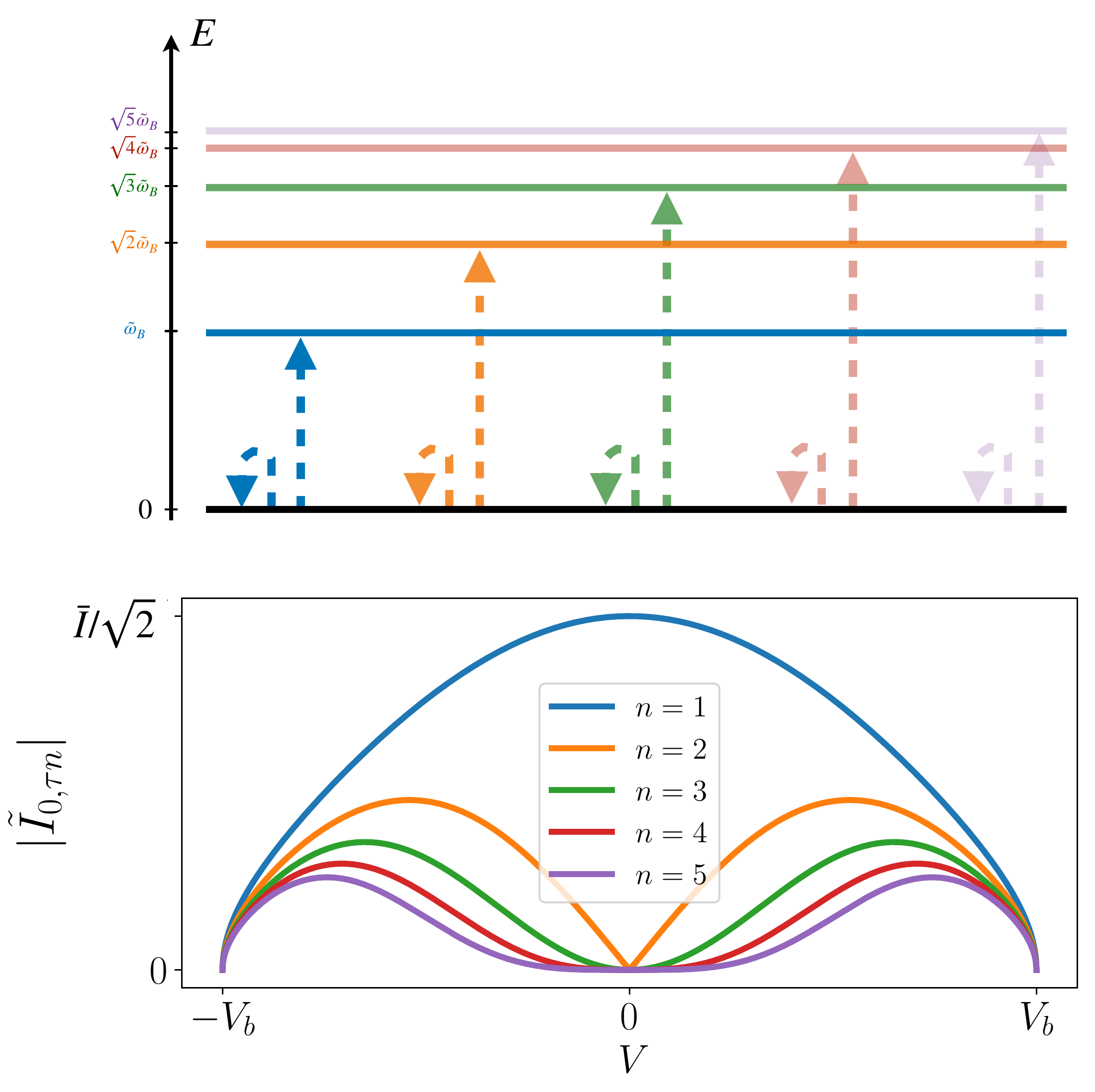}
                \caption{The diagram illustrates a new sequence of allowed transitions involving the parity flip of the Majorana states. At zero voltage bias, only the transition involving the population of the $n=1$ manifold is allowed, while the finite voltage bias induces the transitions to the higher excited states. In the lower panel, we show the corresponding matrix elements as predicted by the formula of Eq. \eqref{majorana_maxi}.}
                \label{majorana transitions}
\end{figure}

The effect of the finite voltage bias is that the Caroli-de Gennes-Matricon (CdGM) states become both squeezed and displaced. Using the raising and lowering operators
\begin{align}
    \hat{a} = \frac{y / \lambda_B + i \lambda_B p_y}{\sqrt{2}}, \quad \hat{a}^\dagger = \frac{y / \lambda_B - i \lambda_B p_y}{\sqrt{2}},
\end{align}
we can express the eigenstates in terms of the number states
\begin{align}
    \ket{\chi_{n, \tau}} = \frac{\ket{\uparrow} \ket{n} + \tau \ket{\downarrow} \ket{n-1}}{\sqrt{2}},
\end{align}
where
\begin{align}
    \ket{n} = \frac{(\hat{a}^\dagger)^n}{n!} \ket{0}, \quad \hat{a} \ket{0} = 0.
\end{align}
Thus, the eigenstates at finite voltage bias are
\begin{align}
    \ket{\tilde{\psi}_0} =& \mathcal{S}\left(\text{arctanh}\left(\frac{\gamma-1}{\gamma+1}\right)\right) \mathcal{U} \ket{\chi_0},\\
    \ket{\tilde{\psi}_{n, \tau}} =& \mathcal{D}\left(\frac{i \beta E_{n, \tau} \lambda_B}{v \sqrt{2}}\right) \mathcal{S}\left(\text{arctanh}\left(\frac{\gamma-1}{\gamma+1}\right)\right) \mathcal{U} \ket{\chi_{n, \tau}},
\end{align}
where \(\mathcal{D}(\alpha) = \exp\left[\alpha \hat{a}^\dagger - \alpha^* \hat{a}\right]\) and \(\mathcal{S}(re^{i\theta}) = \exp\left[\frac{r}{2} (e^{-i\vartheta} \hat{a}^2 - e^{i\vartheta} \hat{a}^{\dagger 2})\right]\) are the displacement and squeeze operators, respectively.

The finite voltage bias renormalizes both the energy and length scales associated with the topological Josephson vortex:
\begin{align}
    \omega_B \rightarrow& \tilde{\omega}_B = \frac{\omega_B}{\gamma^{3/2}} = \left( 1 - \frac{\ell_B^2 V^2}{v^2 \Phi_0^2} \right)^{3/4} \omega_B, \\
    \lambda_B \rightarrow& \tilde{\lambda}_B = \frac{\lambda_B}{\gamma^{1/2}} = \left( 1 - \frac{\ell_B^2 V^2}{v^2 \Phi_0^2} \right)^{1/4} \lambda_B, 
\end{align}
which defines the breakdown voltage
\begin{align}
    V_b = v \frac{\Phi_0}{|\ell_B|} \propto |B|.
\end{align}
In the quasi-relativistic framework, these results can be interpreted as a form of time dilation and length contraction.

\subsection{Current matrix}
Next, we investigate the properties of the current matrix elements, to infer the experimentally accessible (see Refs. [\onlinecite{bretheau2013supercurrent, tosi2019spin, Kurilovich2021, bargerbos2023spectroscopy, wesdorp2024microwave}], for example) transitions between the renormalized CdGM levels.

The current matrix elements are given by:
\begin{align}
\nonumber
    \tilde{I}_{\tau n, \tau' n'} =& \frac{\bar{I}}{2\gamma} \int d\bar{y} \, e^{-i\beta (\tau \sqrt{2n} - \tau' \sqrt{2n'}) \bar{y}} \Bigg( \tau \phi_{n-1}(\bar{y}) \phi_{n'}(\bar{y})\\
    &+ \tau' \phi_n(\bar{y}) \phi_{n'-1}(\bar{y}) \Bigg) \neq 0, \quad \forall n \neq n', \\
\tilde{I}_{0, \tau n} =& \frac{\tau \bar{I}}{i \sqrt{2} \gamma} \int d\bar{y} \, e^{i \tau \beta \sqrt{2n} \bar{y}} \phi_0(\bar{y}) \phi_{n-1}(\bar{y}) \neq 0, \quad \forall n \in \mathbb{N}. 
\end{align}

These expressions show that the application of a finite voltage bias breaks the previously identified selection rule pattern. 

In particular, considering the Majorana transitions, we find:
\begin{align}
    \tilde{I}_{0, \tau n} = \frac{\tau \bar{I}}{i \sqrt{2}} \frac{\left( i \tau \beta \sqrt{n} \right)^{n-1}}{\gamma} \frac{e^{-n \frac{\beta^2}{2}}}{\sqrt{(n-1)!}}, \label{majorana_maxi} 
\end{align}
which has maxima at
\begin{align}
    V_{\text{max}}^2 =V_b^2 \left(1 - \frac{1}{\sqrt{n}} \right)
    \end{align}

Next, we study the current evolution. To begin, we identify the Feynman propagator:
\begin{align}
    K(y, y' | t, t') = \sum_s e^{-i \tilde{E}_s (t - t')} \tilde{\psi}_s(y - \beta v t) \tilde{\psi}_s^\dagger(y' - \beta v t'),
\end{align}
where the sum is taken over all states \(s = 0, (n, \tau)\). This propagator allows us to establish the evolution of the initial (unbiased) excited-state wave functions:
\begin{align}
\nonumber
    \chi_{n, +}(y, t) &= \int dy' K(y, y' | t, 0) \chi_{n, +}(y') \\
    &= \sum_s C_n^s e^{-i \tilde{E}_s t} \tilde{\psi}_s(y - \beta v t), 
\end{align}
with the coefficients:
\begin{align}
    C_n^s = \int dy' \tilde{\psi}_s^\dagger(y') \chi_{n, +}(y'),
\end{align}
defining the effective temperature to which the voltage heats up the system. 

This leads to the expected value of the current (assuming the grand-canonical equilibrium density matrix as the initial state):
\begin{align}
    I(t) = -\bar{I} \sum_{n \in \mathbb{N}} \tanh\left(\frac{E_{n,+}}{2k_B T}\right) \int dy \, \chi_{n,+}^\dagger(y,t) \sigma_x \chi_{n,+}(y,t). 
\end{align}
We can then express this in terms of the matrix elements:
\begin{align}
    I(t) = - \sum_{n \in \mathbb{N}} \tanh\left(\frac{E_{n,+}}{2k_B T}\right) \sum_{s,s'} e^{-i (\tilde{E}_s - \tilde{E}_{s'}) t} (C_n^{s'})^* \tilde{I}_{s',s} C_n^s. 
\end{align}

Finally, averaging the current over time, we obtain:
\begin{align}
    \langle I \rangle = \lim_{\tau \to \infty} \frac{1}{\tau} \int_0^\tau dt \, I(t),
\end{align}
which simplifies to:
\begin{align}
    \langle I \rangle = - \sum_{n \in \mathbb{N}} \tanh\left(\frac{E_{n,+}}{2k_B T}\right) \sum_s \tilde{I}_{s,s} |C_n^s|^2 = 0.
\end{align}
This result indicates that there is no net DC current in the steady-state regime.

\section{Conclusions}
In this study, we analyzed the behavior of Caroli-de Gennes-Matricon (CdGM) states in topological Josephson junctions under the influence of a finite voltage bias. By driving the steady motion of Josephson vortices, the applied voltage significantly alters the system’s spectrum and the localization of CdGM states, introducing a form of spectral squeezing. A critical breakdown voltage was identified, beyond which the states collapse to zero energy and become sharply localized. Crucially, we found that the previously established selection rules governing transitions between CdGM states are modified under voltage bias. Moreover, the steady-state regime exhibits a vanishing time-averaged current, highlighting the delicate balance between vortex dynamics and quantum coherence.

These findings deepen our understanding of nonequilibrium effects in topological superconductors and pave the way for further experimental studies to validate the proposed dynamical phenomena. Future research could explore the role of dissipation, stronger hybridization effects, and interactions between vortices in the voltage-driven regime.

\bibliography{citations}

\end{document}